\documentclass[a4paper]{article}

\usepackage[utf8]{inputenc}
\usepackage[T1]{fontenc}
\usepackage[english]{babel}
\usepackage[bookmarks=false,hidelinks]{hyperref}
\usepackage{lmodern,newunicodechar}
\usepackage[
    backend=bibtex,
    url=false,
    style=trad-plain,
    maxnames=4,
    minnames=3,
    maxbibnames=99,
    giveninits,
    uniquename=init]{biblatex}
\usepackage{microtype}
\usepackage{graphicx}
\usepackage{url}
\usepackage{import}
\usepackage{layout}
\usepackage[table]{xcolor}
\usepackage{amssymb}
\usepackage{amsmath}
\usepackage{csquotes}
\usepackage{enumitem}
\usepackage[binary-units,group-minimum-digits=4]{siunitx}
\usepackage{orcidlink}

\newcommand{\bftab}{\fontseries{b}\selectfont}
\DeclareSIUnit{\million}{\text{million}}

\title{Isabelle as Systems Platform: Managing Automated and Quasi-interactive Builds}
\author{Fabian Huch\thanks{Funded by the Deutsche Forschungsgemeinschaft (DFG, German Research Foundation) under the National Research Data Infrastructure – NFDI 52/1 – 501930651}\;\,\orcidlink{0000-0002-9418-1580}\\ Technische Universität München \\ \texttt{huch@in.tum.de}}
\date{August 2024}

\bibliography{references}

\newcommand{\mizarThms}{$\approx$\;\num{67000} } 
\newcommand{\afpThms}{$\approx$\;\num{271000} } 
\newcommand{\leanThms}{$\approx$\;\num{151800} } 
\newcommand{\afpLoc}{\SI[round-mode=places,round-precision=2]{4.368700}{\million} }
\newcommand{\mizarLoc}{\SI[round-mode=places,round-precision=2]{3.179541}{\million} }
\newcommand{\leanLoc}{\SI[round-mode=places,round-precision=2]{1.562109}{\million} }
\newcommand{\managerSize}{\num{1984} }
\newcommand{\jenkinsDiffDelted}{\num{2810} } 
\newcommand{\scalaModules}{\num{331} }
\newcommand{\scalaLoc}{$\approx$\;\num{63000} } 
\newcommand{\scalaSize}{\SI{3.3}{\mebi\byte} }
\newcommand{\scalaCompressed}{\SI{656}{\kibi\byte} }
\newcommand{\numBuilds}{$\approx$\;\num{21400} } 
\begin{document}
    \maketitle
    \begin{abstract}
Interactive theorem provers are complex systems
that require sophisticated platform efforts --
and hence systems programming environments --
to manage effectively.
The Isabelle platform exemplifies this with its Isabelle/Scala systems programming environment,
which has proven to be very successful.
In contrast, much of the project infrastructure has relied on external tooling in the past, despite shortcomings.
For continuous integration,
the previous system employed a Jenkins server,
which did not adequately support user-submitted Isabelle builds
and faced issues with reliability and performance.
In this work, we present our design and implementation of a new Isabelle build manager
that replaces the old continuous integration system,
fully implemented within Isabelle/Scala.
We illustrate how our implementation utilizes different modules of the environment,
which supported all aspects of the build manager well.
\end{abstract}
    \section{Introduction}
Interactive theorem provers are no longer simple inference engines,
but rather involved systems aimed to support all facets of mathematics.
This has led to the highly sophisticated \emph{prover platforms} we have today.
To build the Isabelle platform,
Isabelle/Scala has emerged as a capable environment for systems programming
after being introduced from 2008\footnote{in Isabelle/606850a6fc1a} to 2010~\cite{Scala2012Wenzel}.
At the time of writing, it consists of \scalaModules modules with \scalaLoc lines of code
that take up \scalaSize (\scalaCompressed compressed).
By providing access to external tools and the operating system uniformly across platforms,
it serves as operating system abstraction,
supporting Isabelle-specific technology, databases,
many file formats, communication channels, and much more.

This approach has been very successful, resulting in high-quality software:
Users are presented with a prover system that simply works out of the box on all platforms.
Moreover, the Archive of Formal Proofs
-- considered the most important outcome of Isabelle prover technology --
is currently the largest uniform body of formalized material in existence
with \afpThms user-specified theorems in \afpLoc lines of code\footnote{\url{https://www.isa-afp.org/statistics}}.
This surpasses both the Lean Mathlib
(\leanThms theorems in \leanLoc lines of code\footnote{\url{https://leanprover-community.github.io/mathlib_stats.html}})
and the Mizar Mathematical Library (\mizarThms theorems in \mizarLoc lines of code\footnote{Source code: \url{http://mizar.org/system/index.html\#download}}),
which was considered the largest body of formalizations in 2009~\cite{Arrow2009Wiedijk}.

In contrast, much of the infrastructure of the Isabelle project --
predating the mature Isabelle/Scala of today --
has been built with external tooling.
In particular, for continuous integration (CI), a customized Jenkins server instance was used
which would re-run all Isabelle sessions as regression test
whenever changes in the underlying repositories occurred.
Moreover, to test changes before publishing them,
users could force-push their commits to a testboard repository,
which would trigger a test run.
Since its introduction in early 2016, the system had performed \numBuilds builds in total.
However, with this many builds (and likely due to software defects),
the Jenkins instance had become quite slow in answering requests, to the point of being unusable.
The system was also prone to leaking memory as well as file handles,
and to spuriously fail starting builds.
Moreover, every job ran on exactly one host drawn from a pre-defined set;
with the advent of cluster-based builds in Isabelle~2024~\cite{Cluster2024Huch},
managing jobs that run on multiple hosts became necessary.
Finally, force-pushing a changeset is not an adequate model for the quasi-interactive Isabelle builds that power users need:
One cannot specify Isabelle build options, the selection of sessions to build, or other parameters.
Also, only changes to one single repository can be tested at a time --
when working with multiple components that need to be adapted and tested simultaneously
(such as Isabelle with the AFP), this is insufficient.

For those reasons, replacing the Jenkins as build manager became necessary.
We decided against utilizing a different existing CI system for several reasons:
\begin{itemize}
    \item Integrating Isabelle properly would require lots of additional work,
    especially since most systems are not designed to handle user-submitted tasks
    and allocate host resources dynamically.
    \item Most of the CI happens inside Isabelle anyway:
    For instance, when AFP entries fail,
    the maintainers of the entry need to be notified by mail.
    \item General-purpose CI systems tend to be quite complex
    since they need to cater to all use cases.
    Maintaining such a system (especially for high compute loads) requires too much effort,
    as our experience with the Jenkins server had shown.
\end{itemize}
Instead, we decided implement the system ourselves within Isabelle/Scala,
exploring whether the platform is mature enough for such a undertaking.
In this work, we describe the \emph{Isabelle build manager} we developed,
and discuss the systems programming aspects.
    \section{System Description}
We implemented the build manager as a single Isabelle/Scala module.
The server is launched via Isabelle/Scala command-line tool
(\texttt{isabelle build\_manager})
and consists of multiple sub-processes,
which run as separate Isabelle threads within the JVM\@.
Similar to the implementation in the build cluster~\cite{Cluster2024Huch},
the state is shared between these (and external) processes via a PostgreSQL database,
using \texttt{isabelle.SQL}.
Hence, the sub-processes can run independently, each performing their own task:
\begin{enumerate}
    \item \emph{Poller}: Listens for repository updates on the Isabelle repository
    and all extra \emph{component}s known to the build manager (such as the AFP).
    When a new changeset for a component appears,
    it queues a \emph{task} for every \emph{CI job} that requires this component.
    \item \emph{Timer}: Queues tasks for CI jobs that should trigger on specific times
    (e.g., nightly at midnight) rather than on every commit.
    \item \emph{Runner}: Checks the queue for tasks whose host selection is currently feasible.
    If such are present, it starts a \emph{job} for the one with highest priority.
    The runner also checks if jobs timed out or were cancelled, interrupting running build processes if needed.
    Processes that don't finish in time after interruption will be terminated forcefully.
    The runner also finalizes any finished job, producing a \emph{result}.
    \item \emph{Web Server}: Provides statically rendered HTML pages of its current state
    when requested (using \texttt{isabelle.HTML} and \texttt{isabelle.Web\_App});
    interactive elements are rendered via HTML 5 forms.
    The state only contains meta-information about build results
    (since there might be tens of thousands of results);
    the build log is read from file when a specific build is accessed.
    Opened log files are cached until no longer requested for a certain time period.
\end{enumerate}

\subsection{Running a Build Processes}
Tasks contain a \emph{build config} that describes either a user-submitted build
(incorporating most parameters of the regular Isabelle build),
or a CI job from the pre-existing \texttt{isabelle.Build\_CI} module.
In either case, running the actual build process consists of four steps:
\begin{enumerate}
    \item The runner uses \texttt{isabelle.Sync} and \texttt{isabelle.Mercurial}
    to create a self-contained copy of Isabelle and the required components
    in the desired revisions, which are given by the task description.
    The revision either refers to the underlying repository of the component,
    or to an already synced file copy (that might contain changes).
    For CI builds, we also store a summary of the commit history
    and changes since the previous build.
    \item Creating the actual build process: This requires initializing the remote
    \texttt{isabelle.Other\_Isabelle} environment for the build via \texttt{isabelle.SSH},
    using \texttt{isabelle.Rsync} to transfer the prepared self-contained copy.
    In contrast to the distributed Isabelle build (where only session directories are considered),
    the extra components are initialized within the remote Isabelle environment.
    \item Running the actual build as a remote bash process (using \texttt{isabelle.Bash}) over SSH\@.
    Importantly, having a handle on the bash process makes it possible to interrupt it
    (e.g., when a user cancels their build).
    Also, the process \texttt{stdout} and \texttt{stderr} are redirected over SSH and written into a log file,
    which can be watched in real time via the web application.
    Since the distributed build requires access to an external build database
    (where only a single build can be active at a time),
    the build manager admits one cluster-based build and any number of single-machine builds in parallel.
    \item When the build is finished, its Isabelle environment is removed and a \emph{report} is written.
    For the latter, use \texttt{isabelle.Build\_Log} to extract meta-information about the build
    (such as the Isabelle revision, start time, etc.), which is then stored in the database.
    While this data could also be directly collected within the Scala process,
    using only the log file has the advantage that all persistent information in the database
    can later be re-constructed from the log files alone.
    This simplifies schema updates, as one can just discard the database
    and re-construct an updated version with \texttt{isabelle build\_manager\_database} --
    dropping temporary information (such as the submitting user and task UUID) in the process.
    Hence, we keep the log files permanently in compressed form.
\end{enumerate}

\subsection{Dynamic Web Application via Static HTML}
The \texttt{isabelle.Web\_App} module inside Isabelle/Scala delivers the static HTML embedded it into an inner \texttt{iframe}.
Hence, if the page contains a form with interactive elements, submitting it
(via any of its buttons)
will trigger a POST request against the server,
and the response will replace the \texttt{iframe} content.
This allows to build interactive web applications from within Isabelle/Scala
while only having to generate static pages,
without requiring clients to execute any JavaScript.
Still, the module delivers some minimal JavaScript
to resize the \texttt{iframe} height to fit the content,
and automatically re-load the content on state changes:
The latter is efficiently achieved by periodically submitting HEAD requests to the server
(which prompt a header-only response)
and checking if the content checksum deviates.

For styling, we \texttt{gd.css}\footnote{\url{https://gdcss.netlify.app/}},
which assigns accessible styles to all visible HTML elements
(and thus does not require any changes in the html structure).

As there is no support for HTTPS in Isabelle/Scala,
we use Caddy\footnote{\url{https://caddyserver.com/}} as a reverse proxy,
which works out of the box with just 4 lines of configuration code.

\subsection{Submitting User Builds}
To make the system as easy to use as possible, submitting build tasks should work very similarly to
performing a local build.
We implemented the \texttt{isabelle build\_task} command,
which has almost the same parameters as \texttt{isabelle build}.
Submitting a build task requires users to connect via SSH to the server where the build manager runs on.
This requirement is the same as for pushing changes to the Isabelle repository.
The \texttt{build\_task} tool then uses \texttt{isabelle.Sync}
to synchronize the local Isabelle/AFP repository clone to the server
(which requires a common unix group, such as the \texttt{isabelle} group on TUM servers),
and opens a connection to the build manager database tunnelled through SSH\@.
If only trustworthy users can access the server via SSH,
the database can be configured a accept local connections without further authentication;
otherwise, users needs to configure Isabelle options for this database connection.
Using the connection, the tool can access the shared state, and register the new task.
Once that is completed, users are given a private URL (containing the randomly assigned UUID of their task),
under which they can cancel it if desired.
A public \emph{name} (which consists of the build kind and a consecutive identifier)
gets created only when the build is started, since the order of queued tasks might change.

\subsection{System Configuration}
\begin{figure}[!htb]
\centering
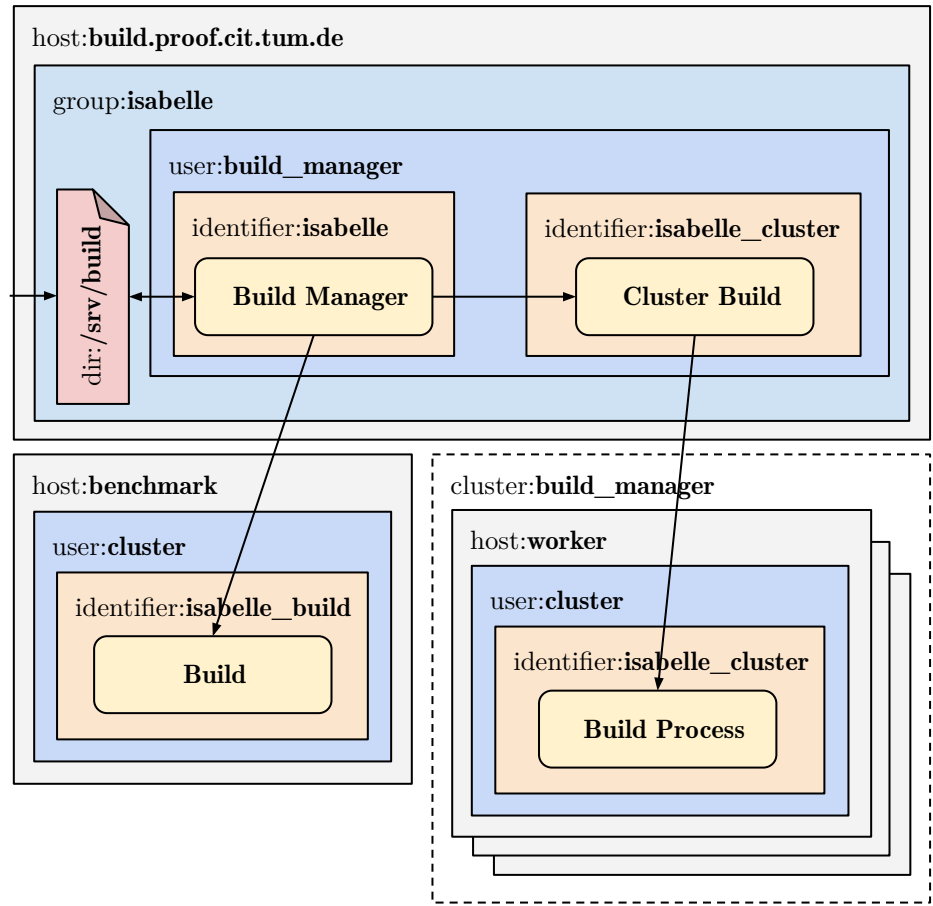
\caption{Isabelle execution contexts for a typical example deployment.}\label{fig:deployment}
\end{figure}

The build manager requires only very little infrastructure,
but several Isabelle execution contexts (i.e., Isabelle user homes with preferences)
are required that need to be configured properly.
Figure~\ref{fig:deployment} shows the involved Isabelle processes and their contexts
for a typical deployment.
The build manager runs on the main host under the common \texttt{isabelle} group,
which also owns the shared directory for user-submitted jobs.
Single-machine builds (e.g., for benchmarking) are run in their own context on remote systems separate from the build cluster,
so they can run undisturbed and in parallel to cluster builds.
In contrast, the master process for a cluster build runs on the main host.
Its Isabelle context is separated from the build manager by using a distinct \texttt{isabelle\_cluster} identifier,
but both processes need similar permissions and secrets so it is reasonable to run them under the same user.
For instance, both require e-mail passwords:
The manager should notify administrators when build errors occur,
and the AFP cluster build has a post-hook to let maintainers know when their sessions fail.
However, it is desirable to run actual build processes (which only need database access) with restricted permissions,
so they are executed under a \texttt{cluster} user which is not in the \texttt{isabelle} group.

\subsection{User Interface}
\begin{figure}[!htb]
\centering
\includegraphics[width=0.99\linewidth]{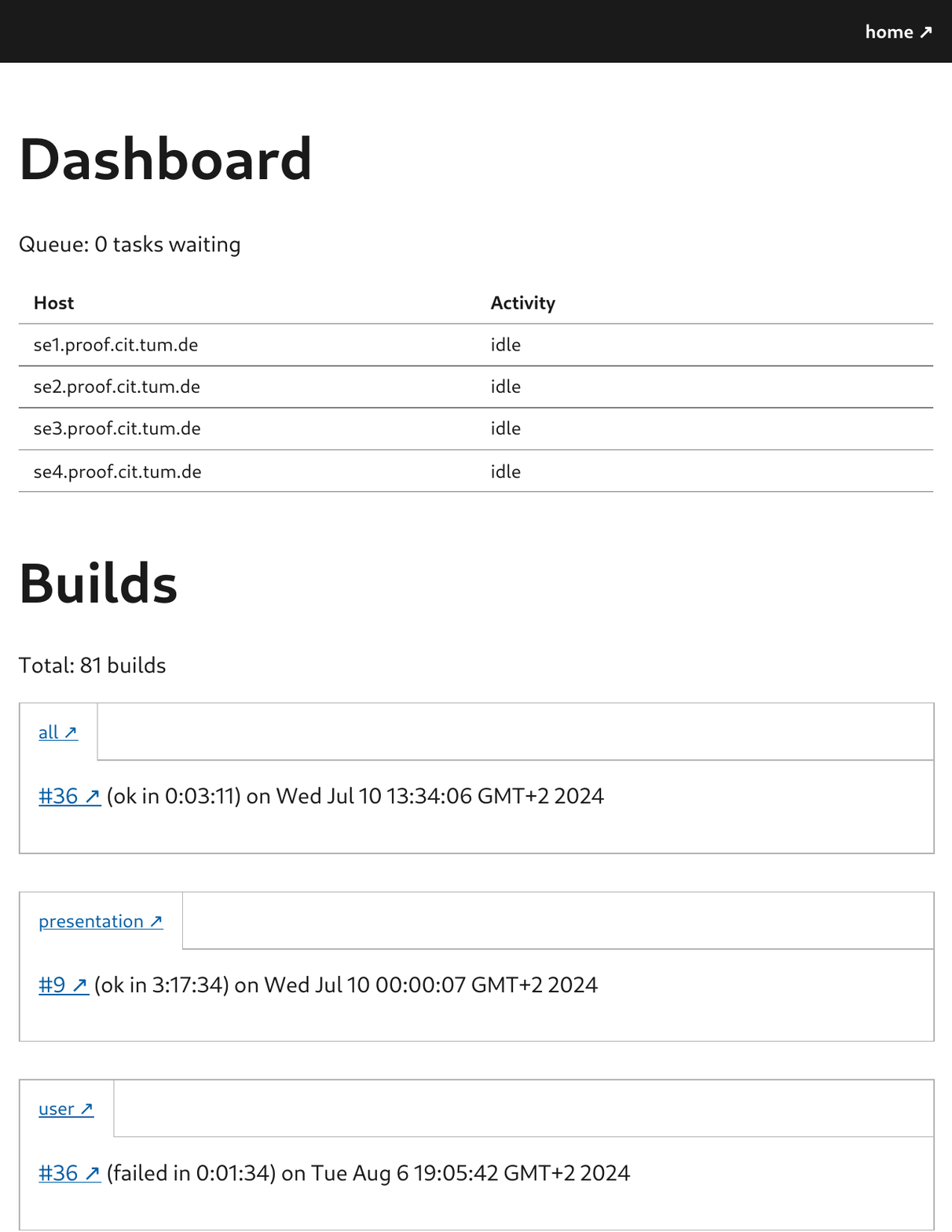}
\caption{Dashboard page of the build manager web application.}\label{fig:frontend}
\end{figure}
\noindent We implemented a minimalistic user interface for the web application,
which is structured similarly to the old Jenkins:
Figure~\ref{fig:frontend} shows the dashboard page,
which summarizes general system statistics and host utilization.
It also gives a short summary for every kind of build,
which contains the link, time, and status of the most recent and other relevant builds.
The summary also links to an overview page,
which lists all builds of its kind.
Finally, the page for an individual build displays meta-information and the build log;
when accessed via private URL, controls to cancel the build are also shown.

    \section{Conclusion and Outlook}
In this paper, we have discussed our design and implementation of the new \texttt{Build\_Manager} module,
which replaces the Jenkins CI server that had been in use since 2016.
We have also highlighted how the module utilizes the rest of the Isabelle/Scala systems programming environment:
With only few changes to the involved modules during the development,
the Isabelle platform could support nearly all aspects of the build manager well --
only HTTPS is handled externally.

The new module contains a total of \managerSize lines of Isabelle/Scala code as of Isabelle/5555a40b2ed4,
replacing \jenkinsDiffDelted lines of Jenkins configuration and integration code in total.
This simplifies the Isabelle environment, and running remote builds is much more accessible than before.
In combination with the speed-up of cluster-based builds
(with a total speed-up factor of over \num{100}),
performing serious Isabelle refactoring is now properly supported again.

Moreover, the new system makes it much easier to adapt managed Isabelle builds to changing technologies:
For instance, many efforts are currently being made to better support research data management
and make research more reproducible,
for instance by standardizing research data management containers~\cite{RMDC2023AlLaban}.
In the future, running such a container could be a standard operation supported by the build manager.

More ambitious goals are now also within reach:
For instance, allowing to update sources live during a running build
(such that emerging problems can be solved in time for other pending sessions)
could further decrease the length of a development cycle.
Also, scaling hosts up and down dynamically might improve hardware utilization even further,
and decrease wait times.
Finally, displaying builds by their output log is merely a minimal implementation:
There are better ways to convey the state of a build,
since one usually wants to know which sessions are currently running (and for how long),
are finished, or have already failed, and how much more work there is to be done.
Similar to the \texttt{Theories} panel in Isabelle/jEdit,
this information could be displayed quite well graphically,
especially with the run time estimation from the \texttt{isabelle.Build\_Schedule} module.
    \printbibliography{}

@inproceedings{Cluster2024Huch,
    author = {Huch, Fabian and Wenzel, Makarius},
    booktitle = {{Interactive Theorem Proving (ITP 2024)}},
    title = {{Distributed Parallel Build for the Isabelle Archive of Formal Proofs}},
    pages = {to appear in},
    series = {{LNCS}},
    publisher = {{Springer}},
    year = {2024}}

@article{Arrow2009Wiedijk,
    title = {{Formalizing Arrow’s theorem}},
    volume = {34},
    ISSN = {0973-7677},
    DOI = {10.1007/s12046-009-0005-1},
    number = {1},
    journal = {Sadhana},
    publisher = {Springer},
    author = {Wiedijk,  Freek},
    year = {2009},
    pages = {193–220}
}

@article{Scala2012Wenzel,
    title = {{Asynchronous Proof Processing with Isabelle/Scala and Isabelle/jEdit}},
    journal = {Electronic Notes in Theoretical Computer Science},
    volume = {285},
    pages = {101-114},
    year = {2012},
    note = {Proceedings of the 9th International Workshop On User Interfaces for Theorem Provers (UITP10)},
    issn = {1571-0661},
    doi = {https://doi.org/10.1016/j.entcs.2012.06.009},
    author = {Makarius Wenzel},
}

@article{RMDC2023AlLaban,
    title = {{Establishing the Research Data Management Container in NFDIxCS}},
    volume = {1},
    ISSN = {2941-296X},
    DOI = {10.52825/cordi.v1i.395},
    journal = {Proceedings of the Conference on Research Data Infrastructure},
    publisher = {TIB Open Publishing},
    author = {Al Laban,  Firas and Bernoth,  Jan and Goedicke,  Michael and Lucke,  Ulrike and Striewe,  Michael and Wieder,  Philipp and Yahyapour,  Ramin},
    year = {2023},
}
\end{document}